\documentclass[letterpaper]{article} 
\usepackage{aaai2026} 
\usepackage{times} 
\usepackage{helvet} 
\usepackage{courier} 
\usepackage[hyphens]{url} 
\usepackage{graphicx} 
\usepackage{natbib}  
\usepackage{threeparttable}
\usepackage{booktabs}
\usepackage{caption}
\usepackage{algorithm}
\usepackage{algorithmic}
\usepackage{tcolorbox}
\usepackage{xcolor,colortbl}
\usepackage{newfloat}
\usepackage{listings}
\DeclareCaptionStyle{ruled}{labelfont=normalfont,labelsep=colon,strut=off} 
\floatstyle{ruled}
\newfloat{listing}{tb}{lst}{}
\floatname{listing}{Listing}

\title{AI-Enabled Human Memory Manipulation:\\Misleading AI-Generated Summaries Distort Human Memory}
\author{
    Mattea Sim\textsuperscript{\rm 1},
    Yael Eiger\textsuperscript{\rm 2},
    Tadayoshi Kohno\textsuperscript{\rm 1}
}
\affiliations{
    \textsuperscript{\rm 1}Georgetown University\\
     \textsuperscript{\rm 2}University of Washington\\

    mattea.sim@georgetown.edu, yeiger@cs.washington.edu, yoshi.kohno@georgetown.edu
}

\begin{document}

\maketitle

\begin{abstract}
AI-generated summaries and reports are increasingly used in high-stakes settings, like policing, despite considerable evidence that AI often generates misleading or inaccurate information. This research asked: do errors in AI-generated summaries distort human memory? To answer this question, we adopted two methodological approaches. First, we conducted an analysis of AI summary output, prompting popular large language models to generate summaries of videos. This analysis quantified how often AI summaries contain errors and the categories of these errors (e.g., omission of critical details; hallucinations), providing insight into the kinds of misleading information that may distort human memory. Second, we conducted a human-subjects experiment to empirically test the impact of misleading information in AI-generated summaries on human memory. Participants were first exposed to an event via watching a video of a car-pedestrian accident, and later read an AI-generated summary describing the video that either contained misleading or accurate information. Participants' memory for the original event was assessed in a memory recognition test. This work contributes several novel findings. In the AI analysis, we found a high frequency of mistakes in AI summaries, and particularly frequent omissions of critical details. For instance, the majority of summaries omitted the most central event of the video, a critical error which is likely to be impactful in high-stakes settings. We also observed a strong effect of AI misinformation on human memory. People who read a misleading AI summary were significantly less likely to accurately recall the original event, compared to people who read an accurate AI summary. These findings have implications for how AI should be used in critical settings. Though ``humans-in-the-loop'' are often expected to correct for AI's mistakes, our work suggests human memory can instead be distorted by these mistakes. AI has the potential to generate misinformation, even absent any adversarial intent, which can meaningfully impact human memory.
\end{abstract}

\begin{links}
    \link{Data and Supplementary Materials}{https://osf.io/wmjtn}
\end{links}

A shortened version of this paper is published in the \textit{Proceedings of the Ninth AAAI/ACM Conference on AI, Ethics, and
Society}, October 2026.

\section{Introduction}

Generative AI is being rapidly adopted and integrated for use in high-stakes settings, like policing~\cite{PoliceLLMs}, hiring~\cite{TheAlgorithm}, and healthcare~\cite{Doximity}. 
For instance, police departments can now automatically generate police reports from body-worn camera footage or audio~\cite{Axon}. However, these AI-generated reports or summaries can include mistakes. In one such example, an AI-generated police report mistakenly claimed a police officer shape-shifted into a frog after picking up on audio from ``The Princess and the Frog'' playing in the background~\cite{Frog}. To account for AI's fallibility in high-stakes settings, guidelines typically suggest human oversight, in which a person should check AI output to ensure the accuracy of the information~\cite{PoliceLLMs}. 

Unfortunately, human memory is also fallible, and the assumption that people can notice and correct for errors in AI output may not always be true. In our work, we ask: how might misleading information in an AI-generated summary impact \textit{human memory}? In fact, human memories can be quite susceptible to mistakes or misleading information. A seminal theory in the field of human memory, the \textit{misinformation effect}, demonstrates that people's memories for events can be distorted by exposure to misleading information or misinformation after an event~\cite{Loftus1978, Loftus2005}.\footnote{Throughout this paper, we use the term ``misinformation'' consistent with its use in the human memory literature, which refers to misinformation via both its definition as misleading or inaccurate information, \textit{and} as the psychological effect misleading information has on human memory (the ``misinformation effect'').} Because AI can relay information (and misinformation) confidently and persuasively~\cite{Danry2025, Kabir24, Kidd2023}, and because these outputs are used in a variety of impactful settings~\cite{PoliceLLMs, Doximity}, there is a real risk of AI creating misinformation at an unprecedented scale --- and perhaps impacting human memory, as well.

We adopted two distinct approaches to explore the impact of AI summaries on human memory. First, we conducted an \textbf{analysis of AI summary output to quantify the frequency and the category of errors that appear in AI-generated summaries.} Here, we prompted popular large language models (LLMs) to summarize videos before quantifying \textit{how often} errors occur and \textit{how} errors occur --- in other words, what categories or classes of errors arise in the output?
Rather than conducting a systematic analysis of a wide range of videos, we conducted a more limited analysis of two brief, straightforward videos that are classically used in the human memory literature~\cite{Loftus1978, GTAStudy} to more deeply understand the AI errors that might impact human memory.
This analysis posed the following hypothesis:
\begin{itemize}
    \item \textbf{H1:} AI-generated summaries will include a meaningful frequency of errors across a range of categories that could be impactful for human memory.
\end{itemize}

Second, we conducted a \textbf{human-subjects experiment to empirically test the impact of errors in AI-generated summaries on human memory}. Here, people witnessed an event (by watching the same classic memory videos used in the above analysis) before reading an AI-generated summary of the event that either contained accurate or misleading information about the event. People then completed a memory test assessing their recollection for the original event. 
Our central hypothesis was:
\begin{itemize}
    \item \textbf{H2:} Misleading information in an AI-generated summary will distort people's memories for an event: i.e., an AI-enabled misinformation effect.
\end{itemize}

We also tested whether people's susceptibility to misinformation would depend on whether they believed the summary was AI-generated versus human-written. To do so, human participants were first told that the summary was written by \textit{either} an AI model \textit{or} a human (though in actuality, the summary was identical across conditions and was generated by ChatGPT). People also responded to questions assessing their trust in AI. This test posed three alternative hypotheses:

\begin{itemize}
    \item \textbf{H3a:} People will be similarly susceptible to misinformation believed to be generated by AI and humans. 
    \item \textbf{H3b:} People will be \textit{more} or \textit{less} susceptible to misinformation believed to be generated by AI than by a human.
    \item \textbf{H3c:} People's susceptibility to misinformation will depend on how much they trust AI.
\end{itemize}

This work makes several contributions. To foreshadow our findings, we observed that AI-generated summaries included a high frequency of errors across a range of categories. For instance, we found frequent omissions of critical details in each AI summary in our analysis, a type of error that can likely degrade human memory. Centrally, we also found that misleading information in an AI-generated summary distorted people's memory. People were susceptible to misleading information in AI summaries, regardless of whether they believed the summary was AI-generated or human-written, and regardless of how much they trusted AI or used AI. These findings have real implications for human-AI collaboration in high-stakes settings. Whereas humans are often positioned to account for and mitigate AI's mistakes, our work demonstrates that these mistakes in AI output could instead manipulate human memory, even for those who might be more skeptical of AI's output. By applying psychological theories of human memory to the rapidly evolving landscape of generative AI, we demonstrate both human memory and AI are fallible, together painting a concerning picture for human-AI collaboration in critical settings where AI is increasingly adopted.

\section{Background and Related Work}

\subsection{Misinformation and False Memories}
For over 50 years, human memory researchers have explored the ways in which our memories are susceptible to manipulation~\cite{Loftus1975, Loftus2005}. This research surfaced a phenomenon now known as \textit{the misinformation effect}: people's recollection of events are often altered after exposure to inaccurate or misleading information~\cite{Loftus1978}. In this classic study (and later replications)~\cite{GTAStudy}, people witnessed a car-pedestrian accident, in which a car comes to an intersection with either a stop sign or a yield sign. After, people were exposed to information about the traffic sign that was either accurate, misleading, or neutral (no mention). Finally, in a forced-choice memory recognition task, people had to recall whether they originally saw a stop sign or a yield sign. Those who were exposed to misleading information were considerably more likely to misremember the critical traffic sign detail. 

The impact of misleading information can be quite powerful. Indeed, even a seemingly minor semantic detail, such as the strength of a verb used to describe a car crash (e.g., ``smashed'' versus ``hit''), can alter later memory for related details like the speed of the cars~\cite{Loftus1974}. More dramatic examples of misinformation include what are sometimes called ``rich false memories,'' in which 25\%--35\% of adult participants were susceptible to a detailed false memory of being lost in the mall as a child~\cite{LostintheMall, LostintheMallAgain}.

Once a false memory is implanted, it can also be quite robust and resilient. The effect of misleading information on human memory often persists even when people are directly warned that they were exposed to misinformation~\cite{Eakin2003}, and when people are told that everything they read after an event was inaccurate~\cite{Lindsay90}. In other words, many people's memories are still swayed by misleading information, even when they are told to ignore the information (cf. Highhouse and Bottrill~\citeyear{Highhouse_Bottrill}).
Once a false memory is implanted, it can be difficult to correct the memory later.

Though the misinformation effect is widely studied, technological advances facilitate novel ways of spreading misinformation and potentially impacting human cognition. In our work, we investigated how AI might facilitate a new iteration of the misinformation effect. 

\subsection{AI Misinformation and Its Impacts on People}

It is now widely documented that LLMs (e.g., ChatGPT, Gemini) have a tendency to ``hallucinate''~\cite{HallucinationCategories, Xu2024}, perpetuate bias~\cite{Gueorguieva26, Nicolas25, raj-etal-2026-talent, Wilson24}, omit relevant information~\cite{omissionsClinical}, and make otherwise inaccurate claims~\cite{Kabir24, Zhou2023}.  Consequently, AI has the potential to spread impactful misinformation, even absent any adversarial intent. 

A full review of the literature on LLM failure rates and categories is beyond the scope of our paper, but this work supports the idea that AI output frequently contains errors that could impact human memory. Researchers have surfaced, quantified, and taxonomized categories of errors in AI-generated summaries~\cite{HallucinationCategories, TaxonomyErrors}. In research on dialogue summaries, for example, the output inaccurately ordered chronological events, missed important details, and made baseless inferences~\cite{InferenceHallucination}. Errors are also studied in higher-stakes contexts where AI-generated summaries are increasingly used. For instance, AI-generated medical summaries sometimes include hallucinations and omissions~\cite{omissionsClinical,ClinicalSummaries}, errors which could be highly impactful in a clinical context. Such mistakes persist in human-AI interactions beyond summarization. For instance, seeking information from ChatGPT often yields inaccurate responses~\cite{Kabir24}.
Further, AI chatbots or companions are prone to highly sycophantic behaviors like flattering and agreeing with users, which can degrade the output's accuracy and reliability~\cite{SycophancyScience, SycophancyAIES, SycophancyInformational}. 

AI video-to-text summaries in particular can also include numerous categories of inaccuracies. Comparing AI-generated video captions across several models to human-written captions revealed AI models generated both hallucinations (broadly defined as inaccurate descriptions or ordering of video details) and omissions of important details~\cite{videosummaries}. This active body of work studies how to quantify and reduce errors in video-to-text models~\cite{videounderstanding, videosums2, videosummaries, Videohallucer}. To date, existing research demonstrates systematic errors in AI video summaries.
Though we do not conduct a large-scale AI analysis to systematically study model behavior, questions which are thoroughly studied by these researchers and many others, our research is the first to begin to categorize AI summary misinformation (i.e., errors) in ways relevant to human memory effects. 

In the face of considerable evidence for the fallibility of AI, increasing attention has turned toward understanding how AI interaction or exposure may be impactful for human cognition. For instance, exposure to biased simulated output from LLMs and text-to-image models can impact humans' own decision-making~\cite{Wilson25} and implicit bias~\cite{Sim2025}. Interactions with AI can even influence personal beliefs and attitudes more powerfully than typical interventions~\cite{persuasion25, science2025}. Indeed, persuasive messages generated by AI are often perceived as stronger than human-generated persuasive messages~\cite{Karinshak2023}. 

Of increasing interest is how AI-generated \textit{misinformation} in particular might impact users~\cite{Kidd2023}. People are often susceptible to this misinformation: for instance, when verifying answers to programming questions provided by ChatGPT, programmers overlooked nearly 40\% of errors in ChatGPT's responses~\cite{Kabir24}. Chatbots can even influence people's political preferences by presenting relevant evidence and facts, including inaccurate and hallucinated claims~\cite{persuasion25, science2025}. Purposefully deceptive misinformation generated by AI can also be highly impactful and persuasive~\cite{Danry2025, goldstein2024, Zhou2023}.
Relevant to the present work, AI misinformation can also impact human memory. Memories for news articles or photos were influenced by interacting with a misleading chatbot~\cite{chatbot_memory25} or exposure to AI-edited images~\cite{CHI25memory}. Building on this work, we study how misleading AI-generated summaries, increasingly used in a variety of contexts, may impact human memory.

A related and novel question investigated in our work is whether people's susceptibility to misinformation is impacted by the belief that the information was human-written versus AI-generated. Information source can impact human perception: people tend to prefer AI-generated information \textit{unless} it is explicitly labeled as AI-generated~\cite{Karinshak2023}. Further, people are aware that AI-generated information contains errors, and thus are more skeptical of its output~\cite{Kabir24}. Yet users also increasingly rely on AI for work, information-gathering, and other high-stakes tasks for which accuracy is key~\cite{Angrisani26, Milella_Cabitza_2026}, suggesting at least some users see AI as a highly reliable source. Further, sources that appear confident and objective, like AI, are more readily trusted~\cite{Kidd2023}. Therefore, we explored whether human memory's susceptibility to misleading information depends on the supposed source: AI or human.

\section{Methodology}
Our methodology consisted of two parts. First, we conducted an analysis of popular AI models with video-to-text summarization capabilities to quantify the number of AI output errors and the category of these errors. Second, we conducted a human-subjects experiment in which we tested the degree to which an error in an AI-generated summary would distort people's memories --- that is, an AI-enabled misinformation effect.

\subsection{AI Summary Analysis Procedure}

We conducted an analysis to quantify and categorize the errors produced in AI-generated video summaries with two of the most popular large language models with video-to-text summarization capabilities: ChatGPT-5.5~\cite{ChatGPT} and Gemini 2.5 Flash-Lite~\cite{Gemini}. We focused the analysis on the two 25-second videos that were also presented to participants in the human-subjects study (see Human Memory Study Procedure below), both of which were used in prior memory studies and depict an animated car-pedestrian accident~\cite{GTAStudy}. Rather than conducting a large-scale analysis of many videos, we conducted a smaller analysis of these two videos, prompting each model five times per video. This analysis allowed us to use the same videos across both the AI analysis and the human study, so that our work is consistent and deeply grounded in the human memory literature~\cite{Loftus1978,GTAStudy}. This also allowed us to investigate the \textit{consistency} of models across time --- that is, do models consistently make the same errors, or do models make different errors randomly across repeated tests? The videos were also a conservative test of errors: videos were short, simplistic, and presented minimal details or events, which may actually \textit{reduce} the likelihood of errors and make any resulting errors more striking. This process led to 20~total video summaries which were qualitatively coded for errors. 

\subsubsection{Videos.}
 The two videos were created in a recent replication study~\cite{GTAStudy}, which modernized materials from a classic human memory study paradigm~\cite{Loftus1978}. We selected these videos because of the significance of this study paradigm in the human memory literature, and because the videos were publicly available online. 
 The videos showed an animated depiction of a car-pedestrian accident created in Grand Theft Auto~V~\cite{GTA}.
 The two videos were similar in content with the exception of a traffic sign: the videos show a red car approaching an intersection with \textit{either} a stop sign or a yield sign. Both videos show the red car turning right at an intersection as a pedestrian walks into the road. The car collides with the pedestrian, who falls to the ground. The pedestrian stands up, the driver of the car exits the vehicle, and the driver and the pedestrian meet.

\subsubsection{Prompting the Models.}
We chose to use a single prompt across the AI analysis because the goal of this work was not to systematically study how different prompts change the output of LLMs, questions that other researchers are studying deeply~\cite{PromptingReview,promptengineering24, PromptReport}. Instead, our goal was to select a high-quality prompt with which we could quantify the different kinds of errors arising in LLM output.

In order to select the prompt, we conducted a series of preliminary tests trying different prompts and models for video summarization. For instance, we experimented with example prompts designed to summarize body cam footage~\cite{bodycamprompt}. We also prompted ChatGPT to generate a prompt idea based on our basic experimental design, which we tested. These prompts all included explicit instructions to maintain neutrality, be factual, and avoid making inferences. When testing prompts that omitted neutrality instructions, we observed what appeared to be qualitatively different summaries as a result, often lower quality with more erroneous assumptions. In order to adopt a conservative approach with reasonable, higher-quality instructions, our prompt included direction to generate neutral and factual summaries. 

Further, we included a word minimum to generate summaries of fairly consistent lengths (at least 300 words), based on some existing guidelines for standard-length AI-generated reports~\cite{AxonNarratives}. We also conservatively added instructions to ensure everything important is included in the summary, due to a high frequency of notable omissions observed in initial testing. These prompt choices were intended to give the model its ``best chance'' at generating accurate summaries. 

The final prompt reads:

\begin{quote}
I am seeking your expertise in generating a narrative from a video. Generate a factual, neutral text summary that is at least 300 words and provides a coherent narrative of everything that happens in this video. Do not introduce new facts, speculation, or interpretations. Make sure everything important is included in the summary. Write in third person, past tense. Produce only the summary text.
\end{quote}

All final queries were done in Spring 2026. To ensure repeated prompt responses were not affected by prior memory of the video, each query was run using a fresh upload of the video and prompt each time, on fresh instantiations of the model, five separate times. This created a corpus of five responses from the yield sign video with Gemini, five responses from the stop sign video with Gemini, five responses from the yield sign video with ChatGPT, five responses from the stop sign video with ChatGPT --- or 20 total summaries; 10 for each video; and 10 for each model. The responses from the model were saved to a CSV. 

\subsubsection{Coding Analysis.}
In order to code the AI summaries, three researchers first independently coded videos for artifacts before meeting to refine and standardize a list of what we considered to be the ``central details'' of the event. The list was generally intended to capture the ``who,'' ``what,'' ``when,'' ``where,'' and ``why'' of the event. The final list included 19 central details.

Prior to coding the summaries, we agreed upon four different classes or categories of errors to analyze based on our initial testing: 1) \textit{omissions}, or failure to mention one or more of the 19~central details; 2) \textit{critical inaccuracies}, or errors describing the 19~central details; 3) \textit{non-critical inaccuracies}, or errors unrelated to the central details; and 4) \textit{additions}, or hallucination-like addition of details not present in the event. Each error could only be coded in a single category.

Two researchers independently coded each resulting summary to quantify and categorize the errors. To do so, both researchers coded each independent occurrence of a central detail omitted, a central detail described inaccurately, a non-central detail described inaccurately, and hallucinated additions, resulting in ``counts'' within each class of error for each summary. Whereas the first two categories were coded based on the pre-determined central details list, the latter two categories were coded without pre-determining a list of non-central details or potential hallucinations; in other words, researchers simply coded any other inaccuracies or additions in the summaries that were inconsistent with the videos. The researchers met to resolve disagreements, reviewing the videos to reach a final consensus.

\subsection{Human Memory Study Procedure}
Participants completed a two-part study online via Prolific. The experimental design was a conceptual replication and extension of a classic memory study paradigm in which participants see a car-pedestrian accident and later read a summary of the accident that manipulates whether the car came to a stop sign or a yield sign, a critical detail which participants are then asked about in a memory recognition task~\cite{Loftus1978,GTAStudy}. We retain these core elements of the paradigm and extend it to investigate whether AI-generated summaries produce a misinformation effect amongst human participants, \textit{and} whether participants' memories are further influenced by believing that the summary is either AI-generated or human-generated. See Appendix for the full materials and measures.

\begin{table*}[tb]
\centering
\setlength{\tabcolsep}{4mm}
\begin{threeparttable}
\begin{tabular}{ll|ll|ll}
\toprule
Gender &  \%    & Age &   \%   & Race/Ethnicity              &  \%   \\ \hline
Man         & 51.8 & 18-24    & 5.2  & White/European American          & 77.1 \\
Woman       & 46.6 & 25-34    & 23.8 & Asian/Asian American             & 10.4 \\
Non-Binary  & 1.2  & 35-44    & 32.3 & Black/African American           & 9.5  \\
            &      & 45-54    & 23.5 & Hispanic/Latino                  & 6.1  \\
            &      & 55-64    & 7.9  & Multiple Identities Selected     & 5.2  \\
            &      & 65+      & 7.3  & Native Hawaiian/Pacific Islander & 0.3  
            \\
\bottomrule
\end{tabular}
\caption{Participant self-reported demographic frequencies.}
\label{tab:demographics}
\end{threeparttable}
\end{table*}

\subsubsection{Part~1: Encoding the Original Event.}
 In the first part of the study, participants were told they would be acting as an eye-witness to an animated car-pedestrian accident and were asked to pay close attention to the video. Participants watched one of two 25-second animated videos created in Grand Theft Auto~V by researchers of the original study; videos are available online via their paper~\cite{GTAStudy}. The videos showed a red car approaching an intersection with \textit{either} a stop sign or a yield sign. The traffic sign was counterbalanced across videos (i.e., people were randomly assigned to see one of the two signs) to rule out potential stimulus effects, such as if one traffic sign was particularly memorable and created unique effects. Otherwise, both videos showed the same events with differences in only peripheral visual details (see description in ``Videos'' section of AI Summary Analysis Procedure). 
 After participants watched the video, they answered an attention check question which asked the color of the main car in the video to ensure they paid attention. Only participants who correctly answered ``red'' were invited to take the second part of the study. 

\subsubsection{Part 2a: Post-Event Summary.}
We implemented a 24--48~hour gap between the two parts of the study, consistent with findings that the impact of misleading information on memory is stronger after a longer time delay~\cite{Loftus1978,GTAStudy}. Twenty-four hours after watching the video, participants who passed the attention check were invited to take part in the rest of the study via Prolific, which was live online for the subsequent 24~hours or until all participants completed the study. 

In the second portion of the experiment, participants read a 21-sentence AI-generated text summary describing the details of the video. 

\textit{\textbf{The AI-Generated Summary.}} In our experimentation with video  summarization, ChatGPT often produced summaries with errors, such as the omission of critical details, and these errors were often irregular or inconsistent across responses. Thus, one researcher (on a separate account from the AI summary analysis prompting) repeatedly re-prompted ChatGPT, ultimately selecting one of the higher-quality summaries returned to show participants (with minor modifications to ensure accuracy, described below; see Appendix for the full original and modified summary).\footnote{After noticing prompts were omitting details of the car-pedestrian accident, one researcher used a new ChatGPT account (separate from the account used in the AI analysis) to re-prompt the model until the accident was included in the resulting summary. Subsequently, in some cases, ChatGPT would return accurate details, whereas in other cases, it would not. We consider it beyond the scope of this study to diagnose exactly when and how ChatGPT can evolve to produce better summaries.} While this re-prompting process may not accurately represent all uses for ChatGPT-based summarization (i.e., we already knew the central details of the video and were striving to ensure the AI summary included those details), our process ensured that the summary used in this part of the study was a valid ChatGPT output.

The final summary was generated using the same prompt as in the AI Summary Analysis Procedure, and summarized the yield-sign video. There were slight peripheral differences between the yield video and the stop video, such as different advertisement boards and surrounding buildings. However, we chose to use one consistent summary for all participants in order to avoid confounding differences across conditions. Indeed, generating two separate summaries for different videos in ChatGPT often yielded meaningfully different text, with variability in descriptions and errors.

Thus, we made only minor edits, such as removing the specific details that were inconsistent across the yield and stop videos so that the summary would be accurate regardless of which video participants watched (e.g., removing descriptions of specific ad signs), and adding minor text that had been previously generated by ChatGPT. For instance, in prior experimentation, AI summaries of the stop sign video described it as a “red stop sign,” and thus, this AI-generated text was injected back into the summary for participants who were assigned to read the stop sign summary. 

\textit{\textbf{Experimental Conditions.}} Prior to reading the summary, participants were first randomly assigned to one of two ``Summary Conditions,'' in which they read that the summary was written by either an AI software designed to professionally transcribe videos (\textit{AI Condition}), or a person who works as a professional video transcriber (\textit{Human Condition}). In actuality, the summary was identical for all participants (with the exception of one critical traffic sign detail) and was generated by ChatGPT. 

In addition, participants were randomly assigned to one of two ``Information Conditions,'' in which the summary included a description of the traffic sign that was either consistent with the sign from the video (\textit{Consistent Condition}) or inconsistent with the sign from the video (\textit{Misleading Condition}) --- in other words, the summary described either a stop sign or a yield sign, either of which could be consistent or misleading depending on the original video. Though we controlled the presence of the traffic sign description to experimentally manipulate misinformation, we did so using actual language output by ChatGPT, so that all text included in the summary was originally AI-generated.

\subsubsection{Part 2b: Remembering the Event.} 
Immediately after reading the summary, participants completed filler questions assessing the quality of the summary. Participants were then instructed to think back to the original video event (\textit{rather than} the AI- or human-generated summary) to answer a series of questions, emphasizing their answers should be based on their own recollection of the video.

Participants responded to a 12-item memory recognition task with two-choice response options based on the details of the video. Ten of these questions were taken directly from prior work~\cite{GTAStudy}. Of central interest was the critical question, asking at which traffic sign the car came to a hold at the intersection. Participants could select either ``stop'' or ``yield.''

Participants responded to a 10-item scale assessing trust in AI~\cite{TrustScale}, and reported their frequency of AI use and other basic demographics before being debriefed. 

\subsubsection{Participants.} 
All participants were recruited from Prolific and compensated at the minimum wage rate of the first author's city, which is higher than the federal minimum wage. We relied on an a priori power analysis conducted in G*Power~\cite{GPower} from prior work~\cite{GTAStudy}, which determined that at least 63 participants per condition would provide 80\% power to detect a medium effect size in a Chi Square test ($w=0.25$, $df=1$). We over-sampled to account for participant drop-off and attention check failures, collecting data from 360~adult U.S.-based participants in part~1. Using Prolific's available screeners, participants were required to be fluent in English and to report having no issues seeing colors. Five of these 360~participants were not invited to participate in the full study due to a failed attention check (but were still paid for their participation in part~1).

In total, 331 participants completed both parts of the study. Three of these participants failed an attention check in part 2, leaving 328 participants included in analyses (AI Consistent Condition $n = 82$; AI Misleading Condition $n = 80$; Human Consistent Condition $n = 83$; Human Misleading Condition $n = 83$). See demographics in Table~\ref{tab:demographics}.

\section{Results}

All data and supplementary appendix materials are available online (\url{https://osf.io/wmjtn}) for reproducibility and open science.

\subsection{AI Summary Analysis Results}
We first quantified and categorized errors in AI summary output (H1). We connected each category of error to findings in the psychology literature that suggest the error could likely be impactful for human memory. 

\subsubsection{AI Summary Errors Are Frequent and Irregular.}
We first tested H1 to quantify the frequency and category of errors in AI-generated summaries of our test videos.\footnote{See Appendix for a table including the counts of each error category occurring across all twenty individual summaries (i.e., across the two models and repeated prompts for the stop and yield videos).}

All twenty summaries included mistakes. Table~\ref{tab:percenterrors} shows the percentage of summaries that included each category of error. We found a \textit{minimum} of seven errors in each summary, up to a maximum of twenty-one errors in a given summary --- notably high in summaries that averaged just 305~words. Both ChatGPT and Gemini made mistakes, though the mistakes themselves descriptively varied. For instance, whereas ChatGPT output included more errors of ``addition,'' Gemini output included more errors when describing central details of the event. In short, both AI models were prone to generating mistakes when summarizing videos. 

\begin{table*}[tb]
\centering
\begin{threeparttable}
\begin{tabular}{l|cccc}
\toprule
        & \multicolumn{1}{l}{Omissions} & \multicolumn{1}{l}{Critical Inaccuracies} & \multicolumn{1}{l}{Non-Critical Inaccuracies} & \multicolumn{1}{l}{Additions} \\ \hline
ChatGPT & 100\%                           & 30\%                                                                   & 80\%                                            & 70\%                            \\
Gemini  & 100\%                           & 100\%                                                                  & 100\%                                           & 50\%                            \\
Total   & 100\%                           & 65\%                                                              & 90\%                                            & 60\%                           \\
\bottomrule
\end{tabular}
\caption{Percent of summaries that included at least one of each category of error.}
\label{tab:percenterrors}
\end{threeparttable}
\end{table*}

The mistakes themselves were often inconsistent across summaries. For instance, a car would be omitted in one summary, but included in the next; hallucinated vehicles or pedestrians would appear spuriously; the main car would be described as turning right in one summary, and left in the next. Although certain errors were particularly common (e.g., omissions described in the next section), overall, errors in AI summaries were not consistent or uniform.

\subsubsection{AI Summaries Omit Critical Details.}
What \textit{kinds} of mistakes arose most often in AI summaries? The most common category of error was omission, or failure to mention the primary events of the video. Unlike the other categories of errors, omission errors were present in every summary. 

Table~\ref{tab:criticalerrors} shows the percentage of \textit{central details} across all summaries that were omitted, inaccurate, or accurate (whereas Table~\ref{tab:percenterrors} shows the percentage of summaries with each category of error, Table~\ref{tab:criticalerrors} shows the percentage of central details that were omitted altogether or misconstrued in some way). Across all prompts and models, the summaries omitted 51.6\% of central details on average. 

\begin{table}[tb]
\begin{threeparttable}
\begin{tabular}{l|lll}
\toprule
   & Omissions & Inaccuracies & Accuracies \\
\hline
ChatGPT & 52.1\%    & 3.2\%         & 44.7\%      \\
Gemini  & 51.1\%     & 9\%           & 40\%        \\
Total   & 51.6\%     & 6.1\%         & 42.4\%     
\\
\bottomrule
\end{tabular}
\caption{Percent of \textit{central details} omitted, inaccurate, or accurate across all summaries in each model.}
\label{tab:criticalerrors}
\end{threeparttable}
\end{table}

Most notably, all but one summary (95\% of summaries) omitted the single most central detail of the event: the car colliding with the pedestrian. 90\% of summaries also omitted related key details of the event, in which the pedestrian is lying on the ground and stands up after the collision. Some summaries even omitted the pedestrian altogether, such as one summary that goes so far as to say ``no other characters or vehicles interacted directly with the scene.''

\begin{tcolorbox}
\textbf{Result:} AI summaries omitted critical details.

\textbf{Connection to Human Memory:} Human memory is often degraded when accurate details are omitted from (human-written) text~\cite{Loftus1978}. 
\end{tcolorbox}

\subsubsection{AI Summaries Misconstrue Critical and Peripheral Details.}
AI summaries also included a notable number of critical inaccuracies, or errors when describing the 19 central details of the videos. 65\% of summaries inaccurately described at least one central detail. Summaries contained an average of 1.15 errors related to central details, or 6.1\% of central details described inaccurately across summaries. For instance, summaries often inaccurately described interactions between different elements of the scene (e.g., a car passing around another car), patterns of movement (e.g., a car turning right, a pedestrian walking into the road), and chronological sequences of events (e.g., a car stopping before a driver exits the car). In our analysis, Gemini appeared more prone to this category of error compared to ChatGPT.

AI summaries misconstrued more peripheral or non-central details of the videos, as well, averaging 2.05 non-critical inaccuracies per summary. For instance, summaries often inaccurately described visual elements of the scene like the text on signs or positions of the cars. 

\begin{tcolorbox}
\textbf{Result:} AI summaries inaccurately described important and minor details.

\textbf{Connection to Human Memory:} Inaccurate descriptions of both central and peripheral details in (human-written) summaries can distort memory~\cite{BalantMisinfo}. 
\end{tcolorbox}

\subsubsection{AI Summaries Hallucinate and Directly Contradict or Alter Central Details.}
Finally, AI summaries also included notable hallucinations or additions of details not included in the videos, with an average of 1.6 errors per summary. Over half of the summaries (60\%) included at least one hallucination or addition. For instance, summaries hallucinated additional pedestrians and vehicles, erratic vehicle movements, and inferences about characters' intentions, all of which meaningfully alter or misconstrue important events in the video. 

Some AI summaries included errors that directly contradicted important details. For instance, one summary stated that the pedestrian slightly repositions but does not enter the roadway, whereas the video shows the pedestrian walking into the road, a key sequence of events which causes the car-pedestrian accident.

\begin{tcolorbox}
\textbf{Result:} AI summaries added erroneous details.

\textbf{Connection to Human Memory:} Inaccurate additions to an event in (human-written) text can implant false memories~\cite{Loftus1975}. 
\end{tcolorbox}

\subsection{Human Memory Study Results}
We next \textit{empirically} tested the claim that an error in an AI summary could distort human memory (H2), with data from the human-subjects experiment. We further investigated whether human susceptibility to misinformation depended on the supposed source of information or people's trust in AI (H3).

\subsubsection{A Misleading AI Summary Distorts Human Memory.}

We tested our central hypothesis (H2) that AI summaries containing misleading information would impair people's memory for that information. There was a significant relationship between the Information Condition and the likelihood of answering the critical traffic sign question correctly, $\chi^2 = 53.937$, $p < .001$. 
Consistent with our hypothesis, 83.6\% of participants who read a consistent summary answered the critical question correctly, compared to only 44.8\% of participants who were correct after reading a misleading summary.\footnote{The misinformation effect was strong in magnitude regardless of whether participants originally saw the stop sign video, $\chi^2 = 37.995$, $p < .001$, $\phi = -.481$, or the yield sign video, $\chi^2 = 25.002$, $p < .001$, $\phi = -.390$, though overall accuracy was higher for those who saw the stop sign video. Prior work found people do encode the traffic sign to memory without summaries, with 80.26\% of people correctly recalling the traffic sign after watching the video~\cite{GTAStudy}.} These results demonstrate a strong AI-enabled misinformation effect, whereby misleading AI summaries distort memory.

\begin{tcolorbox}
\textbf{Memory Finding:} Misleading information in an AI summary can distort people's memory for an event.
\end{tcolorbox}

\subsubsection{People are Similarly Susceptible to AI and Human Misinformation Regardless of Their Trust in AI.}

We next tested whether susceptibility to misleading information would depend on whether people believed the summary was written by an AI model versus a human (H3). There was no significant relationship between the Summary Condition (AI- versus human-labeled) and the likelihood of answering the critical question correctly, within either the consistent condition, $\chi^2 = 1.181$, $p = .277$, $\phi = .085$, or the misleading condition, $\chi^2 = 0.136$, $p = .712$, $\phi = -.029$. 

\begin{figure}[tb]
\centering
\includegraphics{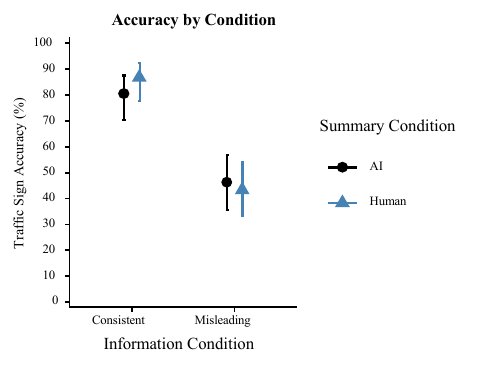}
    \caption{Percent of participants who answered the critical question correctly across conditions. Vertical error bars represent 95\% confidence intervals.}
    \label{figure:accuracybycondition}
\end{figure}

In other words, there was a strong misinformation effect \textit{regardless} of how people believed the summary was written, indicated by significant relationships between the Information Condition and the likelihood of answering the critical question correctly, within both the AI-labeled summary condition, $\chi^2 = 20.499$, $p < .001$, $\phi = -.356$, and the human-labeled summary condition, $\chi^2 = 34.345$, $p < .001$, $\phi = -.455$ (see Figure~\ref{figure:accuracybycondition}).

Descriptively, amongst participants who read misleading summaries, just 46.3\% in the AI condition and 43.4\% in the human condition answered the critical question correctly. Human memory was strongly susceptible to misinformation, regardless of whether the summary was supposedly written by a human or an AI model, supporting Hypothesis 3a.

\begin{tcolorbox}
\textbf{Memory Finding:} 
Misleading AI summaries can consistently distort human memory, regardless of whether summaries are supposedly human-written or AI-written.
\end{tcolorbox}

We also explored whether participants' level of trust in AI or frequency of using AI would moderate the extent to which they were susceptible to misinformation across the AI and human conditions. We first created a composite AI Trust variable averaging across 10 items ($\alpha = .959$; $M = 3.10$, $SD = 0.97$; higher numbers = more AI trust). 

We conducted a binary logistic regression to investigate the interaction between Summary Condition (Human = 0, AI = 1), Information Condition (Consistent = 0, Misleading = 1), and AI Trust (mean-centered) on the likelihood of answering the critical traffic sign question correctly (0 = incorrect; 1 = correct). There were no significant interactions ($ps > .16$), including the central three-way interaction, $\beta = -0.131$, $SE = 0.577$, $p = .820$, $\exp(\beta) = .877$, nor was there a significant relationship between AI trust and the likelihood of answering the critical question correctly, $\beta = -0.344$, $SE = 0.381$, $p = .366$, $\exp(\beta) = .709$. 

Running the same binary logistic model but with participants' reported frequency of AI chatbot use (mean-centered; $M = 3.06$, $SD = 0.93$) replacing AI trust, the three-way interaction was not significant, $\beta = -0.240$, $SE = 0.397$, $p = .547$, $\exp(\beta) = .787$, nor was there a significant relationship between frequency of AI use and the likelihood of answering the critical question correctly, $\beta = -0.068$, $SE = 0.188$, $p = .718$, $\exp(\beta) = .934$. 

\begin{tcolorbox}
    \textbf{Memory Finding:} 
    The extent to which people trust AI or use AI does not change their susceptibility to misleading AI summaries.
\end{tcolorbox}

Therefore, the susceptibility of human memory to misinformation did not depend on whether people believed the summary was written by an AI model or a human, nor did it depend on trust or familiarity with AI.

\subsubsection{Non-Critical Questions.} Finally, we explored several non-critical questions. We tested accuracy on non-critical memory questions to address the possible critique that participants had no memory of the original video event, and were instead relying entirely on the text summary. We tested accuracy on a question which was \textit{not} mentioned in the summary: namely, the time at which the accident occurred (day versus night). 96.6\% of participants correctly identified that the accident occurred in the daytime, providing strong evidence that participants were able to recall the original event. On average, participants answered correctly to 91.4\% out of 11 non-critical questions. 

Although not central to our hypotheses, we also investigated whether participant ratings of summary clarity, readability, or enjoyment would differ across conditions. There were no significant effects of Information Condition ($ps > .410$, $n_p^2s < .003$), nor significant interactions between Summary Condition and Information Condition ($ps > .290$, $n_p^2s < .004$), on any of these variables. However, there was a significant effect of Summary Condition on perceived clarity of the text summary, $F(1,326) = 7.800$, $p = .006$, $n_p^2 = .024$. Participants believed the human-labeled summary was more clearly written ($M = 9.01$, $SD = 1.07$) than the AI-labeled summary ($M = 8.62$, $SD = 1.47$), $t(326) = -2.792$, $p = .006$, $d = -0.31$, $95\% CI [-0.53, -0.09]$ (though in actuality, the summaries were identical). Participants also rated the human-labeled summary as descriptively (but not significantly) more readable ($M = 9.16$, $SD = 1.14$) than the AI-labeled summary ($M = 8.97$, $SD = 1.34$), $p = .159$, $d = -0.16$. 

\section{Discussion}

The extent to which mistakes in AI summaries impact human cognition is critical to understand as AI is increasingly used to summarize important information in real-world contexts. In this work, we adopted two distinct methodological approaches to investigate errors in AI summaries and their impact on human memory.
Centrally, we conducted a human-subjects experiment simulating human-AI interaction in which we empirically demonstrated that misleading information in an AI summary distorted people's memories for an event (H2).  Given the widespread use of LLMs in both organizational contexts and amongst daily individual users~\cite{Angrisani26, Doximity, Milella_Cabitza_2026, PoliceLLMs}, our findings suggest the potential for AI to create misleading information at scale --- with real consequences for human memory.

We also found that people were susceptible to misinformation in AI summaries \textit{regardless} of whether they believed the summary was written by AI or a human (H3). This finding extends our theoretical understanding of the misinformation effect, a phenomenon studied in the psychology literature for over 50~years~\cite{Loftus1975,Loftus2005}. Misleading information, as classically written or administered by humans, has a robust impact on human memory. 
Our findings contribute to a recent wave of work on AI and memory~\cite{chatbot_memory25,CHI25memory}, which extends the misinformation effect to generative AI technologies, and we provide novel evidence that the effect may be robust to the apparent source of misinformation. 
Indeed, people are often more skeptical of information believed to be generated by AI~\cite{Kabir24, Karinshak2023}, though many people also increasingly rely on AI for information~\cite{Angrisani26, Milella_Cabitza_2026}. Despite potentially polarizing opinions about AI, we found that the misinformation effect is strong in magnitude, regardless of whether the source is believed to be AI or human. 

Even further, the impact of misleading AI summaries on human memory did not depend on people's trust or familiarity with AI. These results are particularly striking: even those who are critical of AI may be susceptible to an AI-enabled misinformation effect. In other words, an awareness that AI makes mistakes may not be enough to combat the effect of misleading information in an AI summary on human memory. Yet AI frequently generates mistakes, biases, and hallucinations~\cite{HallucinationCategories, Wilson24, Kabir24}, and in some contexts, people collaborating with AI are expected to notice and mitigate these errors~\cite{Tabassi}. Though we interpret a non-significant statistical finding with caution, our results suggest that solutions like increasing AI literacy alone may not be enough to combat these effects.

\subsection{AI Summary Errors in Human Memory Context}
We also conducted an analysis of AI summary output for videos that are significant to the human memory literature, illustrating a high frequency of errors that are likely to impact human memory (H1). A significant body of work is focused on detecting, categorizing, and minimizing errors in LLM output~\cite{InferenceHallucination, HallucinationCategories,  TaxonomyErrors, Xu2024} and AI video-to-text summaries~\cite{videosums2, videosummaries}. Our findings contribute to this literature and the literature on human-AI interaction by connecting common classes of errors in AI summaries to psychological memory effects, suggesting the errors could have a real impact on people. 

For instance, the most common category of AI summary error was \textit{omission}, in which the summary failed to mention central details of the videos. This type of error has been empirically observed in prior research on AI-generated video summaries~\cite{videosummaries} and AI summaries used in higher-stakes medical settings~\cite{omissionsClinical}. Omission errors also have significant connections to the human memory literature. Some studies of the misinformation effect include a ``no information'' condition, in which people are exposed to summaries or post-event information that omits a critical detail altogether. Though typically considered a control condition, people are less able to accurately recall critical details after exposure to \textit{no information} as compared to accurate information~\cite{Loftus1978}. Therefore, if AI-generated summaries omit important details, this error may have impactful consequences for human memory.

In fact, all categories of errors we observed in AI-generated summaries may be likely to impact human memory. In our human-subjects experiment, we implanted a false detail into an AI summary that is akin to a \textit{critical inaccuracy}, or an inaccurate description of a central detail of the video. This is perhaps one of the most commonly studied types of misinformation in the human memory literature, along with inaccuracies related to more peripheral details~\cite{BalantMisinfo}, with robust effects on human memory (as also demonstrated in our study). Relevant to the error of hallucinations, which are frequently studied in analyses of video-to-text models~\cite{videounderstanding, videosummaries}, other studies demonstrated that ``false presuppositions,'' in which people suggest the existence of objects that were not actually present in an event, can implant false memories of those objects~\cite{Loftus1975}. Thus, hallucinations or erroneous additions in AI summaries may be able to manipulate human memory, as well. 

\subsection{Implications for AI Summaries in Practice}

Broadly, our findings have implications for how AI should be used in real-world settings. As generative AI is increasingly aiding humans in high-impact decisions, most guidelines stress the importance of ``humans-in-the-loop'' to collaborate with AI and ensure the accuracy or appropriateness of final decisions~\cite{Tabassi}. In other words, people are expected to notice and mitigate potential biases or errors in AI output. However, our work and other recent work calls into question the effectiveness of this approach. For instance, when collaborating with AI to make hiring decisions, people tend to over-rely on AI recommendations, even when these recommendations are racially biased~\cite{Wilson25}. 

Our work contributes to growing evidence that the human-in-the-loop strategy is not enough to account for AI's mistakes. That errors in AI-generated summaries distort human memory paints a particularly concerning picture when considering how AI may be used to summarize information in real-world contexts, like policing. Though outlandish mistakes like an officer turning into a frog may be more easily corrected~\cite{Frog}, our work demonstrates that AI summaries could include a range of more subtle (yet still impactful) errors, and these errors can alter human memory. 

Notably, policing has been at the forefront of discussions around the misinformation effect for as long as the effect has been studied in psychology. The misinformation effect is often connected to eyewitness testimony, in which misleading questions or interrogations can manipulate eyewitness memory~\cite{Loftus1975, Loftus2024}. Now, the use of AI in policing and related settings raises new questions about how the misinformation effect might manifest in a legal context. Technological advances can change and exacerbate the influence of misinformation, problems worthy of continued research and deep consideration as AI is adopted into these critical settings.

Finally, although a great deal of work in the area of misinformation studies \textit{deliberate} disinformation campaigns~\cite{goldstein2024,Zhou2023}, in which adversaries may seek to spread misinformation to manipulate or cause harm, the present work highlights a potential harm that could arise \textit{with or without} an adversary. AI generates misleading information without malicious intent, and this misinformation can distort human memory. Together, this presents a real risk of an AI-enabled misinformation effect, whereby AI can widely create misleading information and consequently impact human cognition.

\subsection{Open Questions and Limitations}

The current work had limitations that may motivate future work. First, we conducted a more limited analysis of AI model output, which constrains the conclusions we can make about LLM behavior. This analysis was based on two videos previously utilized in the memory literature depicting video game footage of a simple car accident, which limits the generalizability of this work to other more realistic videos. Further, we used a single standardized prompt and analyzed a smaller number of resulting summaries. The goal of our research was not to systematically study all errors in LLM behavior, but rather to assess and illustrate how errors manifest in videos grounded in the human memory literature. That we found errors with both ChatGPT and Gemini does suggest the possibility for errors could be systemic: in the summaries of two videos, both AI systems had significant issues. These findings are further supplemented by prior research studying a higher volume of longer, more realistic videos across a range of models, which consistently observes similar errors across AI summaries~\cite{videounderstanding, videosums2, videosummaries}. We encourage future research to continue studying realistic videos and summaries to better understand these errors, while also pointing to a significant existing literature that more deeply studies current LLM behavior~\cite{InferenceHallucination, HallucinationCategories,  TaxonomyErrors, Xu2024}.  

Second, we tested whether people's susceptibility to misinformation in a video summary would differ depending on whether they believed the summary was written by AI versus a human. However, in actuality, all participants read an identical summary generated by ChatGPT. This design choice provided us with more experimental control, whereby any differences observed across conditions could be attributed to the manipulation, rather than to confounding differences in summary text. One possible weakness of this design is participants in the human condition could have suspected that the summary was actually AI-generated, reducing potential differences across the AI and human conditions. However, if true, we would not expect to see significant differences in the perceived clarity of the identical summary across human-labeled and AI-labeled conditions. Even further, participants were similarly susceptible to supposedly human or AI misinformation regardless of their trust in AI or how often they reported using AI, suggesting human memory may be similarly fallible regardless of the source.

Third, while video summarization can be a critical task in some contexts, some real-world use cases may summarize other forms of media, like audio transcripts~\cite{Axon}. It will be important for future work to investigate how different forms of media may impact the accuracy and reliability of AI summaries. In fact, one hypothesis is that summaries created from audio transcripts alone could be \textit{less} accurate, given the wealth of visual contextual information that would be missing from such transcripts. The methodological choice to summarize videos may have served as a more conservative test of the errors in AI summary output, but future work could study context-dependent methods of AI summarization. 

Fourth, we conducted a highly controlled human-subjects experiment to expose people to an event and subsequent misinformation, following typical procedures in the psychological memory literature. However, complementary future work may benefit from exploring how the misinformation effect manifests in more naturalistic studies conducted within real high-stakes settings. Further, the conclusions of our work are limited by studying only one type of misinformation in simple animated car accident videos. Future iterations of these studies could include a range of frequently occurring errors (e.g., omissions of critical details) based on more realistic videos or events to further understand the impact of AI errors on human cognition.

Inspired by our findings, another question to explore in future work is whether or how to intervene during human-AI collaboration. If AI summaries are used in high-stakes settings like policing, are there strategies that can be implemented to help officers resist the influence of misleading information? For instance, noticing or detecting discrepancies between the original event and later misinformation can reduce the impact of misleading information on human memory~\cite{Putnam}. Warning people about potential discrepancies or misinformation prior to reading AI-generated summaries may help prompt people to notice errors~\cite{Greene, Loftus2024}. However, warnings are not always effective~\cite{Loftus2005}, and human memory can still be susceptible to misinformation after such warnings~\cite{Eakin2003}. Given the robust effect of misleading information on human memory in our findings, and the critical decisions at stake, a deeper question to consider is whether AI should be used in these contexts at all.

\subsection{Conclusion}
Taken together, we found that errors in AI-generated summaries can distort human memory. The present work contributes to a growing body of research suggesting AI has real consequences for human cognition.
Further, these findings have implications for how AI should be used in high-stakes settings, where humans who are meant to correct for AI's mistakes might instead be susceptible to false memories.

\section*{Ethical Statement}
The study was approved by the human subjects review boards (IRBs) of all authors' institutions. Participants were informed that the full purpose of the research would be provided after completing the study; participants were then debriefed after completing the study. We took additional precautions. For example, while our final video selection was driven by videos used in prior studies of human memory, in our initial (unrestricted) exploration of possible videos, we found (and then chose to exclude) videos with graphic content, such as real police body cam footage. 

\section*{Researcher Positionality Statement}
The authors' backgrounds influence the way we approach these research questions. We are a cross-disciplinary research team with backgrounds in AI ethics, computer security and privacy, and social psychology.

\section*{Acknowledgments}
We are grateful to the reviewers for their feedback to improve the paper. This research was supported by the Robert L. McDevitt, K.S.G., K.C.H.S.\ and Catherine H. McDevitt L.C.H.S.\ Chair in Computer Science at Georgetown University and by the U.S.\ National Science Foundation under Award \#2205171.

\appendix
\section{Appendix}
This Appendix includes the full study materials from the human subjects study, additional information about generating AI summaries, and supplementary results.

\section{Human-Subjects Materials and Measures}
Below are full instructions, materials, and measures shown to participants in the human-subjects study.

\subsection{Study Part 1}

\begin{table*}[t]
\centering
\begin{threeparttable}
\begin{tabular}{r|rrrr|r}
\toprule
       \multicolumn{1}{l}{Summary}         & \multicolumn{1}{l}{Omissions} & \multicolumn{1}{l}{Critical Inaccuracies} & \multicolumn{1}{l}{Non-Critical Inaccuracies} & \multicolumn{1}{l}{Additions} & \multicolumn{1}{l}{Total Errors} \\ \hline
Gemini-Stop-1   & 9                             & 2                                         & 1                                             & 0                             & 12                               \\
Gemini-Stop-2   & 9                             & 2                                         & 3                                             & 1                             & 15                               \\
Gemini-Stop-3   & 10                            & 1                                         & 2                                             & 0                             & 13                               \\
Gemini-Stop-4   & 6                             & 1                                         & 2                                             & 6                             & 15                               \\
Gemini-Stop-5   & 9                             & 1                                         & 2                                             & 0                             & 12                               \\
Gemini-Yield-1  & 7                             & 3                                         & 1                                             & 1                             & 12                               \\
Gemini-Yield-2  & 11                            & 0                                         & 2                                             & 0                             & 13                               \\
Gemini-Yield-3  & 12                            & 0                                         & 5                                             & 3                             & 20                               \\
Gemini-Yield-4  & 10                            & 2                                         & 1                                             & 0                             & 13                               \\
Gemini-Yield-5  & 8                             & 0                                         & 0                                             & 1                             & 9                                \\
ChatGPT-Stop-1  & 8                             & 4                                         & 4                                             & 3                             & 19                               \\
ChatGPT-Stop-2  & 13                            & 1                                         & 3                                             & 1                             & 18                               \\
ChatGPT-Stop-3  & 11                            & 2                                         & 3                                             & 5                             & 21                               \\
ChatGPT-Stop-4  & 10                            & 1                                         & 4                                             & 0                             & 15                               \\
ChatGPT-Stop-5  & 12                            & 2                                         & 4                                             & 0                             & 18                               \\
ChatGPT-Yield-1 & 10                            & 1                                         & 1                                             & 3                             & 15                               \\
ChatGPT-Yield-2 & 6                             & 0                                         & 1                                             & 0                             & 7                                \\
ChatGPT-Yield-3 & 13                            & 0                                         & 0                                             & 1                             & 14                               \\
ChatGPT-Yield-4 & 10                            & 0                                         & 1                                             & 1                             & 12                               \\
ChatGPT-Yield-5 & 12                            & 0                                         & 1                                             & 6                             & 19                              \\
\bottomrule                  
\end{tabular}
\caption{Counts of each category of error across all individual summaries.}
\label{tab:rawcounts}
\end{threeparttable}
\end{table*}

\subsubsection{Instructions.}

\begin{quote}This is a two-part study (you are paid separately for each part). In the first part of the study, you will watch a video of an animated car-pedestrian accident and will act as an eyewitness to the accident. 

Tomorrow, we will invite those who pay close attention to this study to participate in the second part of the study via Prolific. In the second part of this study, you will answer questions about the video you watch today. Please complete the second session as soon as you can as the study will only be available for a maximum of 24 hours. If you only complete the first session, you will be compensated for the study as indicated in the payment. However, if you complete the second part of the study, you will receive an additional \$2.40. 

When you continue, you will watch a short animated video that will start automatically. Please attentively observe the video as you will be asked questions about the video later. You do not need audio. Please watch the video closely.
\end{quote}

\subsubsection{Videos.} \textit{Participants are randomly assigned to watch either the yield-sign video or the stop-sign video, both of which were created by a separate group of researchers and are available online via their paper~\cite{GTAStudy}.}

\subsubsection{Attention Check.}

\begin{itemize}
    \item What color was the main car in the video? (Red; White; Green; Blue; Yellow; Gray; Brown; I don't know)
    \item Have you ever seen this video before today? (Yes; No; Not sure)
\end{itemize}

\subsection{Study Part 2}
\textit{Twenty-four hours after part 1 of the study, participants who passed the attention check (selected ``red'' for the main car color) and indicated they had never seen the video before were invited to take part 2 of the study. Part 2 was available online for the subsequent 24 hours, or until enough participants completed the study.}

\subsubsection{Instructions.}

\begin{quote} This is a two-part study. In the first part of the study, you watched a video of an animated car-pedestrian accident and you were asked to act as an eyewitness to the accident.
Now, you will answer a series of questions about the video that you watched in part 1 of this study. First, you will read a summary describing the video you watched in part 1.
\end{quote}

\subsubsection{Summary Conditions.}
\textit{Participants were randomly assigned to either the AI-labeled summary condition or the human-labeled summary condition.}

\begin{quote} When you continue, you will read a summary that was written by 
\begin{itemize}
    \item AI software designed to professionally transcribe videos [\textit{or}]
    \item a person who works as a professional video transcriber
\end{itemize}
\end{quote}

\subsubsection{Video Summary.} \textit{All participants read the following summary with one sentence presented per page, and a timer of 1 second on each page before participants could proceed to the next page. Participants were randomly assigned to read a summary describing either a stop sign or a yield sign, which served as the Information Condition. That is, either sign could be misleading or consistent information, depending on the original video participants watched.}

\begin{quote} The video opened with a wide view of a rural roadside intersection in a desert-like environment with hills in the background. Utility poles and power lines ran along the road, and sparse vegetation and rocks were visible along the roadside. On the side of the frame stood a tall roadside sign. A [\textbf{red stop sign} / \textbf{triangular yield sign}] was positioned at the corner of the road. A red four-door sedan entered the frame from the bottom center and drove forward toward the intersection. The red car continued forward, slowing slightly as it approached the corner. The camera perspective gradually moved closer toward the intersection as the red car advanced. The red car turned right at the corner near the [\textbf{stop sign} / \textbf{yield sign}] and proceeded onto a straight road that stretched into the distance. On the right side of this road, a man wearing a yellow shirt and dark pants was walking along the roadside near the edge of the pavement. As the red car continued along the road, it moved toward the man. The vehicle then struck the pedestrian. The impact caused the man to fall onto the roadway beside the car. His body came to rest on the pavement near the right side of the vehicle. The red sedan came to a stop shortly after the collision. After the car stopped, the driver exited the vehicle. The driver appeared to be dressed in dark clothing. The pedestrian got up from the road and stood near the front-right area of the car. The driver walked around the vehicle toward the pedestrian. The two individuals approached one another and met in the roadway next to the stopped red car. They stood facing each other near the middle of the lane while the surrounding road remained otherwise empty. The video ended with the two people standing close together beside the red sedan on the quiet roadside.
\end{quote}

\subsubsection{Attention Check and Filler Questions.}

\begin{itemize}
   \item The color test is simple, when asked to enter a color you must enter the word teal in the text box below. Based on the previous instruction, what color have you been asked to enter? [\textit{text entry box}]
    \item Rate the clarity of the written summary (\textit{0 = Very Unclear; 10 = Very Clear}) 
    \item Rate the readability of the written summary (\textit{0 = Very Unreadable; 10 = Very Readable})
    \item Rate your enjoyment of the written summary (\textit{0 = Very Unenjoyable; 10 = Very Enjoyable})
\end{itemize}

\subsubsection{Memory Recognition Task.}

\begin{quote} Although you read a summary written by [\textbf{an AI that transcribes videos} / \textbf{a person who transcribes videos}], we now want you to think back to the video you saw yesterday to answer the following questions. On the following pages, respond to the questions \textbf{based on your own recollection of the original video}. 
\end{quote}

\textit{The order of questions and the order of response options were randomized. Here, the correct answer is listed first, with the exception of the critical traffic sign question (for which the correct answer depends on which video participants were randomly assigned to watch).}

\begin{itemize}
    \item What was the color of the car that hit the pedestrian? (Red; White)
    \item At which traffic sign did the car come to a hold at the intersection? (Stop; Yield)
    \item Which time of day did the accident happen? (Day; Night)
    \item Which direction did the car turn before hitting the pedestrian? (Right; Left)
    \item From which direction did the pedestrian appear before the accident happened? (Right; Left)
    \item What was the terrain of the background environment? (Hilly; Flat)
    \item What was the weather condition when the accident happened? (Dry; Rainy)
    \item How many pedestrians were visible? (One; None)
    \item How many street lights were visible? (Two; Three)
    \item How many people got out of the car to help the pedestrian after the collision? (One; None)
    \item What was the color of the shirt of the pedestrian that was hit? (Yellow; Blue)
    \item What was the color of the shirt of the driver? (Black; Orange)
\end{itemize}

\section{Summary Generation for Human Study}

Here, we present more information on how the video summary was generated to be presented to participants in the human-subjects study. 

As noted in the main text, the summary was generated using the same prompt from the AI Summary Analysis Procedure. A researcher (on a separate account) re-prompted ChatGPT with this prompt and the yield sign video repeatedly, ultimately selecting one of the more overall accurate summaries returned. 

We chose a higher-quality summary of the yield sign video generated by ChatGPT and removed specific sentences describing details that are inconsistent across the yield and stop videos, so that the summary would be largely accurate regardless of which video participant watched. 

We include here the original, full version of the AI-generated summary, with italics and bold text to denote the following human-made changes: 
\begin{itemize}
    \item Text in italics was \textit{removed} from the final summary due to its inconsistency across videos or due to  errors in the descriptions. 
    \item Text in bold was AI-generated text appearing in other summaries that we injected back into the present summary.
\end{itemize}

To expand, we added certain AI-generated text back into the present summary for various reasons. First, the yellow shirt worn by the pedestrian appeared in the majority of other ChatGPT summaries, and was thus injected back into the summary to be more consistent with questions in the memory recognition test (but was still also consistent with the majority of AI-generated summaries, which described the pedestrian wearing a yellow shirt). Second, summaries of the stop sign video described it as a ``red stop sign,'' and thus, this AI-generated text was injected back into the summary for participants who were assigned to read the stop sign summary.

The full summary is below:

\begin{quote} The video opened with a wide view of a rural roadside intersection in a desert-like environment with hills in the background. Utility poles and power lines ran along the road, and sparse vegetation and rocks were visible along the roadside. On the left side of the frame stood a tall roadside sign \textit{listing several businesses, including ``Pills,'' ``Suburban,'' ``Taco Farmer,'' and ``Animal Ark''}. \textit{A paved parking area was visible on the right, where a motorcycle was parked near the curb. A person could be seen walking on the far-right side near the buildings. In the distance across the intersection stood a small building with a sign reading ``Liquor Market.''} A triangular yield sign [/\textbf{red stop sign}] was positioned at the corner of the road. 

A red four-door sedan entered the frame from the bottom center and drove forward toward the intersection. \textit{Shortly afterward, a black vehicle drove past from left to right across the intersection.} The red car continued forward, slowing slightly as it approached the corner. The camera perspective gradually moved closer toward the intersection as the red car advanced. 

The red car turned right at the corner near the yield sign [/\textbf{stop sign}] and proceeded onto a straight road that stretched into the distance. On the right side of this road, a man wearing a \textit{light-colored} [\textbf{yellow}] shirt and dark pants was walking along the roadside near the edge of the pavement. As the red car continued along the road, it moved toward the man. 

The vehicle then struck the pedestrian. The impact caused the man to fall onto the roadway beside the car. His body came to rest on the pavement near the right side of the vehicle. The red sedan came to a stop shortly after the collision. 

After the car stopped, the driver exited the vehicle. The driver appeared to be dressed in dark clothing. The pedestrian got up from the road and stood near the front-right area of the car. The driver walked around the vehicle toward the pedestrian. 

The two individuals approached one another and met in the roadway next to the stopped red car. They stood facing each other near the middle of the lane while the surrounding road remained otherwise empty. The video ended with the two people standing close together beside the red sedan on the quiet roadside.
\end{quote}

\section{Additional AI Summary Analysis Results}
See Table~\ref{tab:rawcounts} for the counts of each error category occurring across all twenty individual summaries (i.e., across the two models and repeated prompts for the stop and yield videos) in the AI summary output analysis.

\end{document}